\documentclass[fleqn,usenatbib]{mnras}

\usepackage{newtxtext,newtxmath}
\usepackage[T1]{fontenc}

\DeclareRobustCommand{\VAN}[3]{#2}
\let\VANthebibliography\thebibliography
\def\thebibliography{\DeclareRobustCommand{\VAN}[3]{##3}\VANthebibliography}

\usepackage{graphicx}	% Including figure files
\usepackage{amsmath}	% Advanced maths commands

\usepackage{amsmath}
\usepackage{color}
\usepackage{orcidlink}

\definecolor{forestgreen}{RGB}{34,139,34}

\newcounter{qnumber}

\title[First TESS free-floating planet candidate]{Searching for Free-Floating Planets with TESS: A few words of clarification}

\author[Kunimoto \& DeRocco et al.]{
Michelle Kunimoto \orcidlink{0000-0001-9269-8060} ,$^{1}$\thanks{E-mail: mkuni@mit.edu}
William DeRocco \orcidlink{0000-0003-1827-9399},$^{2}$
Nolan Smyth \orcidlink{0000-0002-8454-3015},$^{2}$
and Steve Bryson \orcidlink{0000-0003-0081-1797}$^{3}$
\\
$^{1}$Department of Physics and Kavli Institute for Astrophysics and Space Research, Massachusetts Institute of Technology, 77 Massachusetts Avenue,\\ Cambridge, MA 02139, USA\\
$^{2}$Santa Cruz Institute for Particle Physics, University of California, Santa Cruz, CA 95062, USA\\
$^{3}$NASA Ames Research Center, Moffett Field, CA 94035, USA
}

\pubyear{2024}

\begin{document}
\label{firstpage}
\pagerange{\pageref{firstpage}--\pageref{lastpage}}
\maketitle

\begin{abstract}
We recently described the results of an initial search through TESS Sector 61 for free-floating planets. In this short note, we provide important context for our results and clarify the language used in our initial manuscript to ensure that our intended message is appropriately conveyed.
\end{abstract}

% \begin{keywords}
% planets and satellites: detection -- gravitational lensing: micro -- planets and satellites: terrestrial planets
% \end{keywords}

%%%%%%%%%%%%%%%%%%%%%%%%%%%%%%%%%%%%%%%%%%%%%%%%%%

%%%%%%%%%%%%%%%%% BODY OF PAPER %%%%%%%%%%%%%%%%%%

\section{Introduction} \label{sec:intro}

In our recent paper \citet{2024arXiv240411666K}, we described the identification of a first free-floating planet candidate in TESS data. We were excited to share this result and demonstrate that such signals appear and are detectable within existing TESS data. Since posting a preprint of our results on arXiv, we have received many useful comments, and we thank members of the community for their feedback and interest in our work. In particular, we appreciate the recent preprints by \citet{Mroz2024} and \citet{Yang2024} for contributing to the dialogue surrounding our event, and more generally microlensing event detection with TESS.

Our review of this feedback has suggested to us that our intended message was not conveyed appropriately. We feel that it is important to clarify a few points in this short note. The points discussed in this note will be expanded upon in an upcoming revised manuscript. In the following, we will address our choice of language when describing the event we discovered, the possibility of non-FFP interpretations of the event, and the expected yield of FFPs in TESS. We hope the following discussion will provide much needed clarification on our intended message when announcing our result.

\section{Comments}

\begin{enumerate}
    \item \textbf{Use of the word ``candidate:''} We used the word ``candidate'' throughout our paper, adopting the ``innocent until proven guilty'' \citep{2015ApJS..217...31M}  criterion for candidacy established by Kepler and the exoplanet community, applied when the data for a detected event cannot rule out an FFP interpretation. This criterion is \textit{not} based on the intrinsic rarity of an FFP, but solely the morphology of the event itself. However, we have subsequently realized that there may be a difference in convention between our interpretation of the word ``candidate'' and how the microlensing community identifies promising potential microlensing signals, which takes into consideration the expected event rate. We will therefore be amending our manuscript to reflect this distinction, and would like to stress that \textbf{we do not claim to have discovered an FFP}. This level of claim appears to be how our detection has largely been interpreted both by members of the community and by popular media. Rather, \textbf{we detected an event that is consistent with an FFP at the level of the light curve}, which is a critical distinction. We recognize that our original choice of language did not successfully reflect this distinction and could therefore be easily misinterpreted as a claim of FFP detection. We wish to make absolutely clear that this was not our intention, and will be amending our manuscript accordingly.

    \item \textbf{Alternate explanations:} In a recent follow-up to our initial result, \citet{Mroz2024} performed a search through OGLE observations of the source star associated with our event and found evidence for starspots, concluding that the star may be magnetically active and that our event is therefore likely to be a flare. We are completely open to the possibility that our event is due to a flare, and are interested in quantifying the relative likelihood of a flare compared to an FFP. We are currently estimating this likelihood for our revised manuscript, considering factors such as the expected fraction of giant stars that flare, the number of giant star flares expected over TESS Sector 61, the typical energies of such flares, and the rate of symmetric flares compared to classical asymmetric flares. The occurrence rate of symmetric flares has not been estimated in the literature to our knowledge.
    
    On this note, we believe that the challenge of distinguishing between symmetric flares and bona fide microlensing events as well as quantifying their relative likelihoods is an important motivation for a TESS-based search for FFPs. Even if subsequent analyses or observations of our target ultimately indicate that our event is a flare (or heartbeat star, or any other non-microlensing scenario), TESS still provides an important opportunity to better understand short-duration false positives in space-based microlensing searches. As such, there is considerable value in a timely TESS search since these explorations will support the analysis and interpretation of FFP candidates with Roman and Earth 2.0.

    \item \textbf{Low expected yield:} In our initial manuscript, we performed simple order of magnitude estimates of event rates and yields to motivate performing a search through existing TESS data for FFPs. We estimated that a search of all currently observed sectors would yield $\sim \mathcal{O}(1)$ FFP. \citet{Yang2024} have since performed a significantly more detailed analysis of potential FFP detection rates in TESS, and also found that the expected yield is low, on the order of one event after 7 years of the TESS mission based on a fainter dataset. We fully acknowledge and agree that the TESS FFP yield will be low, and are excited for Roman and Earth 2.0 to find thousands of FFPs; TESS will not compete with that number.  
    
    However, the low FFP yield is not an indication that a TESS FFP search is poorly motivated. The fiducial power-law used by our work and \citet{Yang2024} to estimate the underlying abundance of FFPs has large uncertainties \citep{sumi2023freefloating}, especially in the low-mass regime. The abundance of FFPs at sub-terrestrial masses is largely unconstrained, and TESS provides an \textit{immediate} opportunity to either make an initial estimate of the abundance, or to place initial constraints on this population. In either case, TESS's results will narrow the uncertainty on the mass function of FFPs in the sub-terrestrial range, which will only benefit dedicated upcoming surveys such as Roman and Earth 2.0.

\end{enumerate}

\section{Conclusion}

We hope that the above discussion has both clarified our initial intent when sharing our detection and has made clear why we believe TESS is a unique opportunity to learn more about space-based FFP detection at a critical juncture prior to the launch of Roman and Earth 2.0. Our detection from Sector 61 has shown that TESS is capable of finding signals consistent with microlensing by low-mass FFPs, at the level of the data. Regardless of the precise nature of our event, TESS will significantly improve our understanding of false positive signals in space-based microlensing searches as well as place new constraints on the sub-terrestrial-mass FFP yield. Our revised manuscript will include significant changes to the language we use in discussing our event, a quantitative analysis of alternate explanations (such as highly-energetic symmetric stellar flares), and a larger discussion regarding our motivation and excitement for a TESS FFP search especially in support of Roman and Earth 2.0. We deeply appreciate the feedback we have received so far from the community and will make sure it is reflected in our revised manuscript as well as future papers on our ongoing search.



\bibliographystyle{mnras}
\bibliography{sample631} % if your bibtex file is called example.bib

%\begin{thebibliography}{99}
%\end{thebibliography}

%%%%%%%%%%%%%%%%%%%%%%%%%%%%%%%%%%%%%%%%%%%%%%%%%%

%%%%%%%%%%%%%%%%% APPENDICES %%%%%%%%%%%%%%%%%%%%%

% \appendix

% \section{Some extra material}

% If you want to present additional material which would interrupt the flow of the main paper,
% it can be placed in an Appendix which appears after the list of references.

%%%%%%%%%%%%%%%%%%%%%%%%%%%%%%%%%%%%%%%%%%%%%%%%%%

% Don't change these lines
\bsp	% typesetting comment
\label{lastpage}
\end{document}